\documentclass[12pt]{article}
\usepackage[a4paper, total={6in, 9in}]{geometry}
\usepackage{authblk}
\usepackage{newtxtext}
\usepackage{newtxmath}
\usepackage{natbib}
\usepackage{hyperref}
\usepackage{booktabs}
\usepackage{enumitem}
\usepackage{mathtools}
\usepackage{graphicx}
\usepackage{physics}
\usepackage[dvipsnames]{xcolor}
\hypersetup{
	colorlinks = true,
	urlcolor   = blue,
	citecolor  = blue,
}

\everymath{\displaystyle}

\newcommand{\eref}[1]{(\ref{#1})}

\newcommand{\ci}{{\tilde{C}}}
\graphicspath{{figs/}}
\title{Higher order moments of scalar within a plume in a turbulent boundary layer}
\author[1,2,*]{M. Pang}
\author[3]{K. M. Talluru}
\author[1]{K. Chauhan}
\affil[1]{Centre for Wind, Waves, and Water, School of Civil Engineering, The University of Sydney, Sydney, NSW 2006, Australia}
\affil[2]{Laboratory of Building and Environmental Aerodynamics, Institute for Water and Environment (IWU), Karlsruhe Institute of Technology, 76133 Karlsruhe, Germany}
\affil[3]{School of Engineering and Technology, University of New South Wales Canberra, Campbell, ACT 2612, Australia}
\affil[*]{miaoyan.pang@kit.edu}
\date{}
\begin{document}
\maketitle
\begin{abstract}
This study examines the statistical nature of instantaneous scalar concentration in an elevated point-source plume (neutral or buoyant) dispersing within a turbulent boundary layer. Using high-frequency long-duration experimental measurements, we extensively validate the gamma distribution as the appropriate probability density function of concentration, particularly at large scalar magnitudes. The two-parameter gamma distribution is shown to capture the PDF at all locations across the plume. The classical similarity of the mean and root-mean-square (RMS) concentration, often expressed through a Gaussian form, is recovered through similarity of the scale and shape parameters of the gamma distribution. In addition,  statistics of extreme events, such as the 99th percentile of the instantaneous concentration signal, are also well predicted, and their observed invariance near the plume centreline is reasoned. Further, similarity is observed for the third- and higher-order central moments and standardised central moments from the experimental data. The framework of the gamma distribution is also analytically extended to higher-order statistics. The experimental data are in good agreement with the predicted central moments up to the eighth order. The results emphasise the importance of achieving statistical convergence for the intermittent concentration signal, directly influenced by finite sampling times in a measurement. A secondary result is obtained for the ratio of plume half-widths based on the mean and the RMS concentration to be $1/\sqrt{2}$, consistent with experimental observations. The results establish the gamma distribution as a consistent and unified model for all scalar concentration statistics in elevated point source plumes within a turbulent boundary layer.
\end{abstract}

\section{Introduction}
\label{sec.intro}
\par The dispersion of a scalar plume within turbulent boundary layers is a classical problem relevant to environmental and engineering applications, including atmospheric pollution, thermal effluents, and industrial mixing. Considerable effort has been devoted to characterising mean and root-mean-square (RMS) concentrations, and these primary statistics are now relatively well characterised even in complex flow scenarios. However, the inherently multi-scale nature of turbulence, coupled with the intermittent and positively skewed nature of scalar fluctuations, thwarts the reliable prediction of instantaneous concentration fields downstream of a source. This limitation is particularly severe when considering high-concentration events, which, although relatively infrequent, are of practical significance, for example, in assessing pollutant exposure and risk. To address this challenge, statistical approaches based on probability density functions (PDFs) of concentration are utilised, as they represent all statistical moments and predict the likelihood of extreme events. Although several distributions have been proposed, a unified framework capable of accurately reproducing not only the primary (mean and RMS) statistics but also higher-order moments has remained elusive, particularly with experimental validation. The present study aims to demonstrate that the gamma distribution offers such a framework, providing a consistent description of scalar concentration PDFs and enabling reliable prediction of statistics, with validation against experimental data along the way.
\par In general, the instantaneous concentration field, $\tilde{C}(x,y,z,,t)$, is decomposed into mean and fluctuating components as $\tilde{C}(x,y,z,,t) = C(x,y,z,) + c(x,y,z,,t)$. Here, $x$, $y$, and $z$ denote the streamwise, spanwise, and vertical directions, respectively. For an elevated point-source plume, where the plume width is much smaller than the boundary layer thickness, the vertical profiles of mean concentration, $C$ and its standard deviation, $\sigma_c$, are well-described by the Gaussian plume model \citep{Fackrell1982, Nironi2015_P1, Talluru2017_scalarMeas}. At a particular streamwise distance from the source, and along the symmetry plane of the plume, the Gaussian model is then expressed as,

\begin{align}
	\frac{C(z)}{C_{0}} &= \exp\left[-(\ln2)\;\xi_C^2\right], \;  \xi_C=\frac{z-z_\text{0}}{\delta_C},
	\label{eq.CmeanGauss}\\
	\frac{\sigma_c(z)}{\sigma_{c,0}} &= \exp\left[-(\ln2)\;\xi_\sigma^2\right], \;  \xi_\sigma=\frac{z-z_\text{0}}{\delta_\sigma}.\label{eq.CrmsGauss}
\end{align}
Further, $z_\text{0}$ is the height of the centreline of the plume (typically the source height for non-buoyant elevated sources), $C_0$ and $\sigma_{c,0}$ are the mean and standard deviation of the concentration at $z_\text{0}$, respectively, and $\delta_C$ and $\delta_\sigma$ are the corresponding half-widths inferred from the mean and standard deviation profiles, respectively. The half-widths are such that $C(z = z_0 \pm \delta_C) = 0.5 C_0$ and $\sigma_{c}(z = z_0 \pm \delta_\sigma) = 0.5\sigma_{c,0}$. Many past experimental studies showed good agreement with this model \cite[see, for example,][]{Fackrell1982, Nironi2015_P1, talluru2018_uc}. The same model can also describe the variance of concentration fluctuations from an elevated source.
For third- and fourth-order moments, it has been shown by many that the spanwise profiles of skewness ($\tilde{C}_\text{sk}$) and kurtosis ($\tilde{C}_\text{ku}$) exhibit self-similarity \citep[see][for example]{Chatwin1990_CCrms, Sawford1992_heatLine, Nironi2013_thesis}. Furthermore, there exists a relationship between $\tilde{C}_\text{sk}$ and $\tilde{C}_\text{ku}$, given by:
\begin{equation}
	\tilde{C}_\text{ku}=A(\tilde{C}_\text{sk}^2)+B,
	\label{eq.KuSk}
\end{equation}
where $A$ and $B$ are constants. \cite{Schopflocher2005_SkKu} claim that these constants may vary depending on the specific flow, spatial, and temporal conditions; experimental and field data suggest that such variations are typically small.
\par Several approaches have been developed for predicting higher-order concentration moments. \cite{Chatwin1990_CCrms} established a method relating $\overline{c^n}$ to $C$ using a two-parameter polynomial and a constant, where $\overline{c^n}$ denotes the $n$th central moment of $\tilde{C}$. Two subsequent studies extended this framework: \cite{Sawford_1995_moments} related $\overline{c^n}$, to $C$ but required an additional parameter for each successive moment. Their parameters are downstream distance-dependent and require extensive parameterisation. \cite{Mole1995_CHighMoments} concluded that relationships between higher-order moments could be parameter-free (see Equation 10 in their paper). An alternative approach involves solving the Cauchy problem of the transport equation \citep{Lebedev2004_CauchyTransportEqn}. Using this method, \cite{2011_Skvortsov_cn_waterchannel} proposed a scaling law relating higher-order moments to concentration variance. More recently, \cite{Bisignano2017_cn} derived expressions for $\overline{c^n}$ using the Fluctuating Plume Model. Despite these theoretical advances, experimental validation has been limited: most studies verified predictions only up to 4$^\text{th}$-order moments, with \cite{Mole1995_CHighMoments} extending validation to 5$^\text{th}$ order.
\par Since the probability density function (PDF) encapsulates all statistical moments, the above observations point naturally to the possibility of self-similar concentration PDFs. Thus, a variety of functional forms have been proposed in the past to represent concentration PDFs, including Weibull, beta, gamma, log-normal, and normal distributions \cite[summarised in table 2 in][]{Cassiani2020}. Among these, the gamma distribution is found to provide better agreement with experimental data \citep{Lung2002_gamma, Duplat2008, Yee2011_gammaPDF, Nironi2015_P1, Efthimiou2016}. The gamma distribution is defined as,
\begin{equation}
	\mathcal{P}(\tilde{C}) = \frac{1}{\Gamma(k)\theta^{k}}\,\tilde{C}^{k-1}\,\exp\left( -\frac{\tilde{C}}{\theta} \right),
	\label{eq.gamma}
\end{equation}
where $k$ is the shape parameter, $\Gamma(k)$ is the gamma function of $k$, and $\theta$ is the scale parameter. 
\par The tail of PDFs is of great significance because higher pollutant concentrations pose greater environmental concerns, whereas measurements at low concentrations are constrained by instrumental detection limits \citep{Cassiani2020}. High-percentile statistics, such as the 90\textsuperscript{th} or 99\textsuperscript{th} percentiles, are therefore commonly employed in regulatory contexts to quantify odour exposure \citep[e.g.][]{Invernizzi2020_OdourHour}. From a theoretical perspective, \citet{Pumir1991_PDF_expTail} proposed a phenomenological model in which a mean passive scalar gradient in the plume produces exponential tails in the scalar PDF. In practice, however, characterising concentration tails remains challenging, as reliable convergence requires prohibitively long sampling durations. Only a limited number of studies have attempted to fit functional forms with specific attention to the PDF tails, with approaches ranging from generalised Pareto to Weibull distributions, and with considerable variation in the extent of agreement reported \citep{Munro2001_PDFtail, Schopflocher2001_PDFtailGPD, Mole2008, Efthimiou2016, Oettl2017}.
\par The representative functional form of concentration PDFs and their tail characteristics lacks consensus in the literature. Laboratory studies are often hindered by slow-response instrumentation or limited sampling periods. The present study addresses these limitations by analysing a comprehensive dataset from recent plume experiments \citep{Pang2023_thesis,Pang2024_plume_exp}, enabling a detailed assessment of concentration PDFs.
\par The paper is organised as follows. $\S$\ref{sec.exp} describes the experimental configuration and datasets used for validation. In $\S$\ref{sec.PDF}, previously proposed PDF models are compared, with particular attention given to the parameterisation of the gamma distribution. $\S$\ref{sec.datacfgamma} assesses the statistical behaviour, also reported in earlier studies, compared with predictions based on the gamma distribution framework. Finally, $\S$\ref{sec.cn} describes the model for predicting higher-order moments, verified up to eighth order, using the gamma distribution.
\section{Experimental setup and data}
\label{sec.exp}
The analysis and results presented in this paper are from wind-tunnel measurements of neutral and buoyant point-source plumes dispersed within a turbulent boundary layer \citep{Pang2023_thesis}. These experiments are conducted in a closed-circuit, large-scale boundary layer facility at the University of Sydney. A detailed account of the experimental design and methodology can be found in \citet{Pang2024_plume_exp}; here, only the essential features and parameters relevant to this paper are summarised.
\begin{figure}
	\centering
	\includegraphics[trim = 100 180 120 200, clip, width =\textwidth]{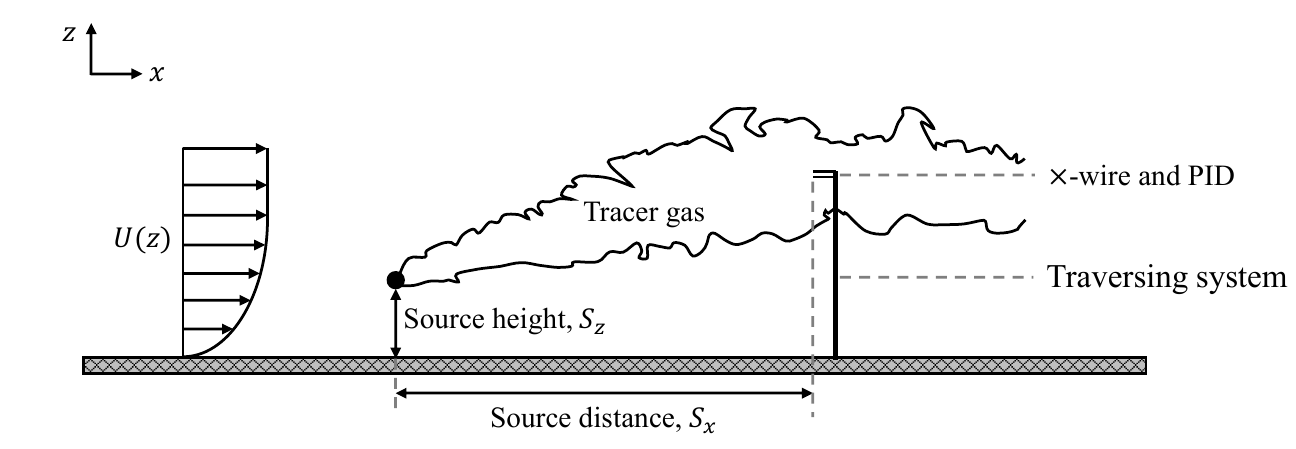} 
	\caption{A schematic of the experimental setup used to study the dispersion of neutral and buoyant scalar plumes in a turbulent boundary layer.}
	\label{fig.schematics}
\end{figure}
\begin{table}
	\begin{center}
		\begin{tabular}{p{1.1cm}p{1.1cm}p{1.1cm}p{1.6cm}p{2cm}}
			\toprule
			$S_z/\delta$ & $\rho_s/\rho_\infty$ & $u_s$, m/s & $Ri$ & $S_x/\delta$:\,Legend \\ \midrule
			0.16 & 0.17 & 1.6 & 9.93\texttimes 10\textsuperscript{-2} & 1:\,{\color{Red}{$\medcirc$}}, 2:\,{\color{Red}{$\bigtriangledown$}}, 4:\,{\color{Red}{{$\triangle$}}}\\
			0.16 & 1 & 1.6 & 6.58\texttimes 10\textsuperscript{-5} & 1:\,{\color{ForestGreen}{$\medcirc$}}, 2:\,{\color{ForestGreen}{$\bigtriangledown$}}, 4:\,{\color{ForestGreen}{{$\triangle$}}} \\
			0.16 & 1.47 & 1.6 & 8.14\texttimes 10\textsuperscript{-3} & 1:\,{\color{Blue}{$\medcirc$}}, 2:\,{\color{Blue}{$\bigtriangledown$}}, 4:\,{\color{Blue}{{$\triangle$}}}\vspace{6pt}\\
			0.32 & 0.17 & 1.9 & 6.89\texttimes 10\textsuperscript{-2} & 1:\,{\color{Red}{$\square$}}, 2:\,{\color{Red}\rotatebox{90}{$\triangle$}}, 4:\,{\color{Red}\rotatebox{90}{$\bigtriangledown$}} \\
			0.32 & 1 & 1.9 & 3.72\texttimes 10\textsuperscript{-5} & 1:\,{\color{ForestGreen}{$\square$}}, 2:\,{\color{ForestGreen}\rotatebox{90}{$\triangle$}}, 4:\,{\color{ForestGreen}\rotatebox{90}{$\bigtriangledown$}} \\
			0.32 & 1.47 & 1.9 & 5.66\texttimes 10\textsuperscript{-3} & 1:\,{\color{Blue}{$\square$}}, 2:\,{\color{Blue}\rotatebox{90}{$\triangle$}}, 4:\,{\color{Blue}\rotatebox{90}{$\bigtriangledown$}}\\ \bottomrule
		\end{tabular}
	\end{center}
	\caption{Summary of parameters for the 18 experiments for the point-source concentration measurements at two source heights ($S_z$), with three density ratios ($\rho_s/\rho_\infty$), and at three downstream distances ($S_x$). The source velocity and the bulk Richardson numbers are denoted as $u_s$ and $Ri$, respectively.}
	\label{tb.config}
\end{table}
\par Figure \ref{fig.schematics}  shows a schematic of the experimental setup, where a point scalar is released at different source heights in a turbulent boundary layer. The turbulent boundary layer has a thickness of 0.255 m and a freestream velocity of 2.5 m/s, yielding a Reynolds number $Re_\delta\approx 43100$. The gas mixture consisted of 98.5\% stable gas and 1.5\% iso-butylene. Iso-butylene (a hydrocarbon) serves as the tracer gas that is detectable by a photo-ionisation device (PID). The stable gas, which is either Helium, Nitrogen, or Argon, is used to adjust the overall mixture density and thereby control buoyancy. The gas is released iso-kinetically, i.e. the source velocity of the tracer gas is matched with the mean local velocity at the source height within the turbulent boundary layer.
\par The buoyancy of the gas mixture is quantified by the density ratio, $\rho_s/\rho_\infty$, where $\rho_s$ is the source density and $\rho_\infty$ is the density of ambient air in the wind tunnel. Table \ref{tb.config} summarises the 18 experimental configurations corresponding to two source heights ($S_z$), three source distances ($S_x$), and three density ratios. The bulk Richardson numbers in table \ref{tb.config} are defined as,

\begin{equation*}
	Ri = gd_s\dfrac{|\rho_s-\rho_\infty|}{\rho_\infty u^2_s},
\end{equation*}
where $g$ is the gravitation constant, $d_s$ is the source diameter, and $u_s$ is the source velocity. The Schmidt number for each gas, $Sc$, represents the ratio between the viscosity and the diffusivity of the gas and is a property of the gas instead of the flow. Here $Sc_\text{He}\approx 0.29$,  $Sc_\text{N}\approx 0.77$, and $Sc_\text{Ar}\approx 0.78$, where the subscripts denote Helium, Nitrogen, and Argon, respectively.

\par The PID from Aurora Scientific (mini-PID 200B) was used to measure the instantaneous concentration at various source heights and source distances downstream of the source. The sensor is mounted on an automated traverse to measure concentration in the vertical direction, above and below the plume source height at each downstream location. These experiments were designed to ensure sufficient spatial resolution across the plume, with approximately 30 sampling points for each configuration. The PID voltage output was low-pass filtered at 400\,Hz and digitised using a 24-bit National Instruments NI-4303 analogue-to-digital converter. At each location, signals were acquired for 300\,s. The sensor was calibrated before and after each experimental configuration, and the raw voltage signals were converted to concentration (in parts per million) following the procedure described by \citet{Talluru2017_scalarMeas}.

\par Two aspects distinguish the present measurements: (i) using a fast-response PID with true frequency of 300\,Hz, and (ii) long sampling durations. Given the intermittent nature of the concentration signal, a fast sensor response is essential for reliably detecting consecutive events of interest. Extended sampling time, as implemented here, is necessary to ensure convergence of higher-order statistics and to resolve the tails of the probability density function with confidence. Typically, PDFs near the plume centreline span more than four decades of concentration magnitude on the ordinate.
\begin{figure}
	\centering
	\includegraphics[scale = 1]{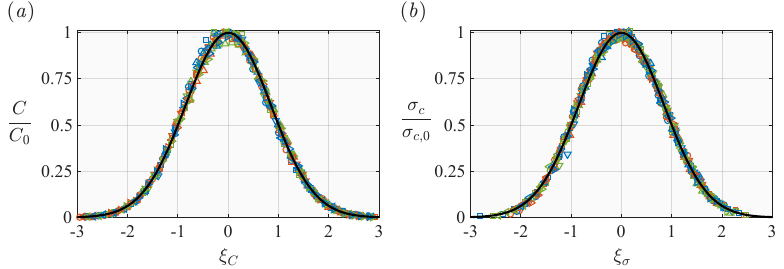}
	\caption{ Normalised profiles of (\textit{a}) mean concentration and (\textit{b}) root-mean-square(RMS) of concentration fluctuations. The black solid lines in (\textit{a}) and (\textit{b}) are the Gaussian model described by equations \eqref{eq.CmeanGauss} and \eqref{eq.CrmsGauss}.}
	\label{fig.GaussianMeanStdDev}
\end{figure}
\par The experimental measurements are first assessed for the concentration mean and root-mean-square (RMS), in figure \ref{fig.GaussianMeanStdDev} (\textit{a} \& \textit{b}), respectively. The mean and RMS of concentration are normalised by the centreline value for comparison with equations \eqref{eq.CmeanGauss} and \eqref{eq.CrmsGauss}. The Gaussian model is in excellent agreement with the data, as also documented in previous studies \citep[e.g.][]{Fackrell1982, Nironi2015_P1, Talluru2017_scalarMeas}. It should be noted that the Gaussian models for $C$ and $\sigma_c$ have different half-widths, i.e. $\delta_C$ and $\delta_\sigma$. Equations \eref{eq.CmeanGauss}  and \eref{eq.CrmsGauss} can be rewritten as,

\begin{equation}
	\frac{C(z)}{C_0} = \left(\frac{\sigma_c(z)}{\sigma_{c,0}}\right)^{\delta_\sigma^2/\delta_C^2}
\end{equation}
Hence, it is evident that $C$ and $\sigma_c^{(\delta_\sigma^2/\delta_C^2)}$ have an invariant relationship across the plume. We later establish that the ratio $\delta_C/\delta_\sigma \approx 1/\sqrt{2}$, as shown in figure \ref{fig.deltaratio}. The relevant discussion is presented in $\S$ \ref{sec.cn} and Appendix \ref{append.delta_ratio}.

\section{Probability density functions}
\label{sec.PDF}
\par Various distributions have been used in the literature \cite[see table 2,][]{Cassiani2020}\ to represent the probability density function (PDF) of concentration measurements at a fixed point. Assessing the suitability of a specific family of probability distributions for experimental data is influenced by several factors, including the fidelity of the experimental data, the relevance of the chosen distribution family, and subjective choices such as the range of fitting and the optimisation algorithm used. In this study, several distributions previously applied to concentration PDFs (listed in table \ref{tb.litrev.PDFs}, Appendix \ref{appen.dist}) were re-evaluated under the same fitting conditions, {i.e.} identical range selection, iterative algorithm, and outlier removal. The fitting was performed in MATLAB using the Curve Fitting Toolbox, with the \texttt{fit} function and \texttt{Method} set to \texttt{NonlinearLeastSquares}.
\begin{figure}
	\centering
	\includegraphics[scale = 1]{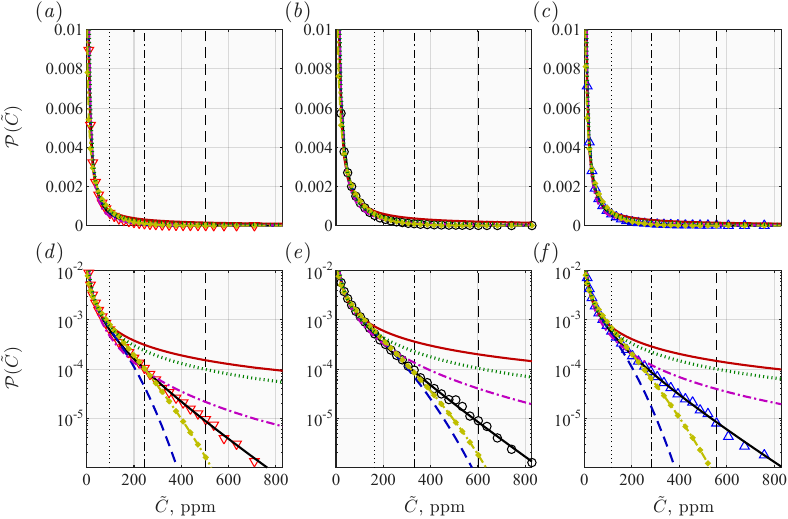}
	\caption{Comparison of distributions for the concentration PDF on log -- linear axes and linear -- linear axes for $S_z/\delta= 0.32$, $S_x/\delta=2$. (\textit{a,d}), (\textit{b,e}), and (\textit{c,f}) correspond to measurements at $\xi_C \approx -1$, $\xi_C \approx 0$, and $\xi_C \approx 1$, respectively. Symbols as per table \ref{tb.config}. Solid black line (\textemdash): gamma distribution, solid red line ({\textcolor{red}{\textemdash}}): log-normal distribution, dashed blue line ({\textcolor{blue}{-\,-}}): normal distribution, dashed yellow line (\textcolor{yellow}{-\,-\,-}): Weibull distribution, dashed-dotted magenta (\textcolor{magenta}{-\,$\cdot$\,-}): beta distribution, and dotted green line (\textcolor{green}{\raisebox{2pt}{\dots}}): generalised Pareto distribution (GPD). (\textit{a-c}) are on log-linear axes, and (\textit{d-f}) are on linear-linear axes. The vertical dotted, dashed-dotted, and dashed lines indicate the 95\textsuperscript{th}, 99\textsuperscript{th}, and 99.9\textsuperscript{th} percentile of the PDF.}
	\label{fig.cf_dist}
\end{figure}
\par Figure \ref{fig.cf_dist} shows the probability density function $\mathcal{P}(\tilde{C})$ for a selected experimental configuration in comparison with the best-fit distributions (listed in table \ref{tb.litrev.PDFs}, Appendix \ref{appen.dist}). Typical concentration PDFs are heavy-tailed and thus in the linear-linear scale in figures \ref{fig.cf_dist} (\textit{a} - \textit{c}), as $\mathcal{P}\rightarrow 0$, the differences between the data and the fitted distributions are not clearly evident. In this representation, all the fitted profiles visually show a good agreement with the data; however, when the same data is plotted on a log-linear scale in figures \ref{fig.cf_dist} (\textit{d} - \textit{f}), the differences become more pronounced for the higher percentiles. The distributions, whose tails do not decay exponentially, over-predict or under-predict at high values. It is observed that up to the 95\textsuperscript{th} percentile, all distributions agree well with the data. The normal, log-normal, and generalised Pareto distributions (GPD) deviate the most beyond this. The beta and Weibull distributions extend their agreement with the data further up to the 99\textsuperscript{th} percentile, but start deviating thereafter. Only the gamma distribution describes well the entire range, even beyond the 99.9\textsuperscript{th} percentile. These observations are consistent with data at other locations within the plume and for other configurations (neutral and buoyant plumes). The gamma distribution is the best fit for the concentration PDF, particularly for the high-concentration tail, which is well-captured in our measurements due to the long sampling time. Based on the above excellent agreement, the analysis presented hereafter considers the gamma distribution as the underlying distribution for the concentration data. The gamma distribution is a two-parameter distribution, and $\S$\,\ref{ssec.premultipliedpdf} outlines the methodology for determining the optimal gamma distribution parameters to fit the experimental data.
\subsection{The gamma distribution fitted to experimental data}
\label{ssec.premultipliedpdf}
\begin{figure}
	\centering
	\includegraphics[width = \textwidth]{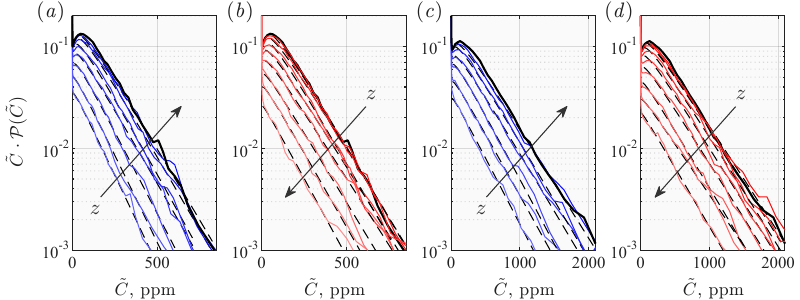}
	\caption{Pre-multiplied probability density functions of the instantaneous concentration, $\tilde{C}$. (\textit{a-b}) $S_z/\delta$ = 0.32, $S_x/\delta$ = 2, $\rho_s/\rho_\infty\approx1$. (\textit{c-d}) $S_z/\delta$ = 0.32, $S_x/\delta$ = 1, $\rho_s/\rho_\infty\approx 0.17$.  The solid black line represents the PDF on the plume centreline, the red lines are PDFs above the centreline, and the blue lines are the PDFs below the centreline. The dashed black lines are the best-fit gamma distributions.}
	\label{fig.cPDF}
\end{figure}
The gamma distribution is characterised by two positive parameters: the shape parameter ($k>0$) and the scale parameter ($\theta>0$). In the limiting cases, $k=1$ reduces the gamma distribution to an exponential form, while it approaches a normal distribution for $k \gg 1$. The scale parameter $\theta$ controls the spread of the distribution, with larger values corresponding to broader profiles and longer tails. For a gamma distribution, the mean value $C$ is given by $k\theta$, and by substituting $\theta = {C}/{k}$ into equation \eqref{eq.gamma}, the pre-multiplied PDF can be expressed as,

\begin{align}
	\tilde{C}\cdot \mathcal{P}(\tilde{C}) &= \frac{1}{\Gamma(k)\theta^{k}}\,\tilde{C}^{k}\,\exp\left( -\frac{\tilde{C}}{\theta} \right) \label{eq.gamma1}\\
	~&= \frac{k^k}{\Gamma(k)} \left[\frac{\tilde{C}}{C}\exp{\left(-\frac{\tilde{C}}{C}\right)}\right]^k.
	\label{eq.gamma2}
\end{align}
Equation \eqref{eq.gamma2} is fitted to the experimental data to estimate the shape parameter $k$. Figure \ref{fig.cPDF} presents the pre-multiplied PDFs obtained from two measurement configurations with markedly different concentration ranges, alongside the corresponding fitted curves for comparison. The high-concentration tails of the PDFs are dominated by the exponential term in equation \eqref{eq.gamma2}, and therefore appear as straight lines when plotted on log–linear axes. The fitted gamma curves reproduce this behaviour well, showing excellent agreement with the experimental data in the tail region. This agreement is consistent across the plume and over a wide range of concentrations. At the low end of the distribution, concentration values close to zero are excluded from the fitting procedure owing to the finite intermittency observed in the measurements, {i.e.} $\mathcal{P}(\tilde{C}=0)\ne0$. Once the shape parameter $k$ is obtained from the fit, the scale parameter $\theta$ can be evaluated using the method of moments.

\par More broadly, equation \eqref{eq.gamma2} provides an excellent representation of the data, not only for the representative configurations shown in figure \ref{fig.cPDF}, but also for all other data analysed in this study. An additional observation is that the high-concentration tails in the pre-multiplied PDFs are nearly parallel. This implies that the scale parameter $\theta$ in equation \eqref{eq.gamma1} remains approximately constant across different cases. The constant slope of tails suggests a statistical similarity in the behaviour of high-concentration events, a point that is explored further in $\S$ \ref{ssec.gammavalidation}.

\subsection{The shape parameter \texorpdfstring{$k$}{k} and the scale parameter \texorpdfstring{$\theta$}{theta}}
\label{ssec.gammavalidation}
\begin{figure}
	\centering
	\includegraphics[scale = 1]{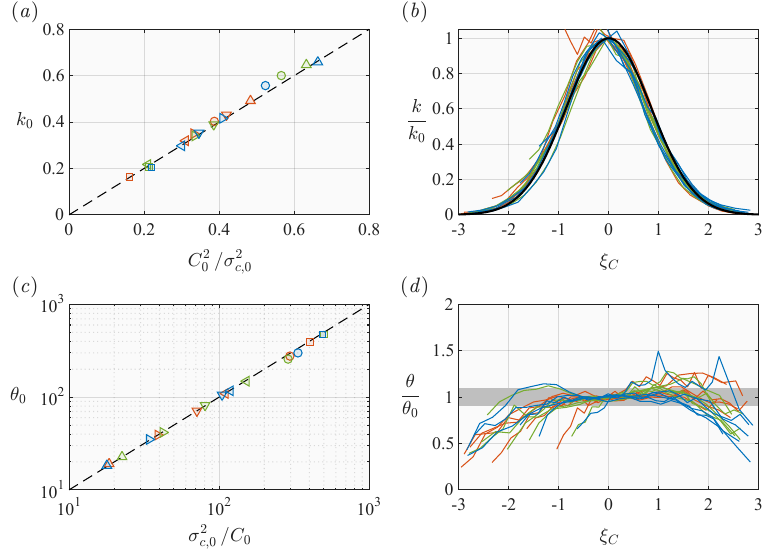}
	\caption{(\textit{a}) Fitted shape parameter for the measurements at the centreline ($k_0$) vs. the statistical prediction of equation \eqref{eq.GammaTheta}. (\textit{b}) Variation of fitted shape parameter $k(z)$ relative to the magnitude at centreline versus normalised distance from the plume centreline $\xi_C$. (\textit{c}) Scale parameter evaluated at the centreline ($\theta_0$) vs. the statistical prediction of equation \eqref{eq.GammaK}. (\textit{d}) Shape parameter, $\theta$, of the gamma distribution vs. normalised distance from the plume centreline $\xi_C$. Symbols as defined in table \ref{tb.config}. }
	\label{fig.GammaFitParameters}
\end{figure}
This section examines the behaviour of the parameters $k$ and $\theta$ obtained from fitting experimental data, extending the scope beyond what has been reported in previous studies. We also compare the behaviour of $k$ and $\theta$ with the theoretical characteristics of a gamma distribution to evaluate its suitability for describing meandering plumes. For the gamma distribution given by $\mathcal{P}(\tilde{C}) = f(k,\theta)$, we can define:

\begin{align}
	&\text{Mean,}\ C = k\theta    \label{eq.GammaMean}\\
	&\text{Standard deviation,}\ \sigma_c = \sqrt{k}\theta \label{eq.GammaCrms}\\
	&\text{Shape parameter,}\ k	= (C/\sigma_c)^2 \label{eq.GammaK}\\
	&\text{Scale parameter,}\ \theta = \sigma_c^2/C \label{eq.GammaTheta}
\end{align}
\par Figure \ref{fig.GammaFitParameters}(\textit{a}) shows the fitted shape parameter $k$ at the centreline, denoted as $k_0$, versus the theoretical prediction of equation \eqref{eq.GammaK}, i.e. $k_0=(C_0/\sigma_{c,0})^2$. The data show a good agreement with the theoretical prediction, indicating that the ratio of mean and standard deviation of concentration on the plume centreline well describes the shape parameter. The fitted $k(z)$ away from the centreline is also plotted against $\xi_C$, in figure \ref{fig.GammaFitParameters}(\textit{b}). The $k(z)$ profiles are observed to be Gaussian-like, and relative to the centreline value, they exhibit a similar variation to the mean concentration profile in equation \eqref{eq.CmeanGauss}. Thus, the shape parameter $k$ can be considered to have a Gaussian similarity in the vertical direction, similar to the mean concentration. The $k(z)/k_0$ profiles in figure \ref{fig.GammaFitParameters}(\textit{b}) agree remarkably well with the right-hand side of equation \eqref{eq.CmeanGauss}. Thus, the shape parameter $k$ can be expressed as a function of the normalised distance from the centreline, $\xi_C$, as:
\begin{equation}
	\frac{k(z)}{k_\text{0}} = \exp\left[-(\ln2)\;\xi_C^2\right], \;  \xi_C=\frac{z-z_\text{0}}{\delta_C}.
	\label{eq.kGauss}
\end{equation}
\par Two important conclusions can be made from figure \ref{fig.GammaFitParameters}(\textit{b}). First, the shape parameter $k$ is not constant across the plume, but varies with the distance from the centreline. Second, the variation of $k$ is similar to that of the mean concentration and described by equation \eqref{eq.kGauss}. It follows then that the shape parameter $k(z)$ at any height can be predicted if $\delta_C$ (the mean concentration half-width) is known, and $k_0$ is known. The latter can be estimated in two ways:
\begin{enumerate}
	\item ~If the time-varying $\tilde{C}(\xi_C = 0)$ is known, $k_0$ is estimated from fitting to the PDF of $\tilde{C}$ at the centreline.
	\item ~Or, if the time-averaged statistics $C_0$ and $\sigma_{c,0}$ are known, $k_0$ is calculated using equation \eqref{eq.GammaK}.
\end{enumerate}
\par This is a significant finding, allowing the prediction of the shape parameter $k$ without additional parameterisation beyond what is required for the prediction of $C(z)$ and $\sigma_{c,_0}$; describing $k(z)$ throughout the plume. Previously, the parameterisation of the downstream evolution of $k_0$ for ground-level sources includes \cite{Duplat2008}, who proposed $k_0\propto x^q$, where $q\approx 5/2$, with \cite{Yee2011_gammaPDF} further suggesting that $q$ depends on the power law coefficient of the velocity profile and eddy diffusivity. Experimental evidence from \cite{Hoot1973_heavy}, \cite{1997_Sykes_plume_secondorder} and \cite{2011_Skvortsov_cn_waterchannel} supports power-law decay of $C_0$ and $\sigma_{c,0}$, which, when combined with equation \eqref{eq.GammaTheta}, yields $k_0\propto x^q$. However, experiments by \cite{karnik1989heatdiffusion, Stapountzis1986_templinesource, Lavertu2005_linsource_channel} show that the decay rates of $C$ and $\sigma_c$ are not constant but vary with downstream distance and background flow statistics, suggesting that $k_0\propto x^q$ is insufficiently comprehensive and that parametrising the exponent $q$ is challenging. Thus, modelling $k_0$, $C$, and $\sigma_c$ for streamwise evolution remains fragmented in previous studies, whose focus was on the centreline shape parameter, $k_0$, providing no information on the vertical and spanwise variation of $k$. This limitation is addressed with the present method; $k(z)$ can be calculated directly from $C(z)$ and $\sigma_{c,0}$, which are available in existing plume models. 
\par The scale parameter $\theta_0$, experimentally determined at the plume centreline is plotted in figure \ref{fig.GammaFitParameters}(\textit{c}) versus the analytical prediction, i.e. $\theta_0 = {\sigma_{c,0}^2}/{C_0}$. Again, the agreement of experimental values with theory is equally consistent for the scale parameter as that for the shape parameter in figure \ref{fig.GammaFitParameters}(\textit{a}). The profiles of $\theta(z)$ relative to the centreline values are plotted in figure \ref{fig.GammaFitParameters}(\textit{d}). As mentioned for figure \ref{fig.cPDF}, the PDF has parallel tails, indicating that the exponentially dominated behaviour at high concentrations could be modelled using a single $\theta$ value at different heights. This is indeed observed in figure \ref{fig.GammaFitParameters}(\textit{d}) that $\theta$ do not vary much from the centreline values, especially closer to the centreline within $-1\lesssim \xi_C \lesssim 1$. Our finding reconciles the conclusions from two independent studies. First, \cite{Mole2008} argued that the tail of $\mathcal{P}(\tilde{C})$ exhibits universality because molecular diffusion and local turbulent structures responsible for concentration transport are uniform across the plume. Second, mathematical analyses by \cite{Pumir1991_PDF_expTail} and \cite{1994_Shraiman_PDF_exponentialTail} demonstrated that $\mathcal{P}(\tilde{C})$ must exhibit exponential tails when large-scale mean scalar gradients are present.
The observation of a constant scale (or rate of decay) of concentration tail is also interesting to note in conjunction with the similarity of concentration spectra observed by \cite{Talluru2019_Spectra}, which will be discussed further in $\S$\ref{sec.discussion}.
\par The tail of the PDF near the edge of the plume ($|\xi_C| \gtrsim 2$) on either side has a different decay exponent than the centreline values. This indicates a decreased statistical convergence towards the plume edge \citep{Chatwin1990_CCrms, 1997Yee_ClippedGammaDeduct}. However, for simplicity, one can assume $\theta(z)$ to be invariant across the whole plume for a simplified characterisation of PDF using a gamma distribution, $\Gamma(k,\theta_0)$, where $k(z)$ is from equation \eqref{eq.kGauss}. Under this assumption, $\theta$ varies with the streamwise distance as the plume evolves spatially. Although outside the bounds of $|\xi_C|\gtrsim2$, the predicted statistics with an invariant $\theta$ assumption are still in good agreement with the data, and are further presented later in $\S$\ref{sec.cn}.
\section{Comparison with known statistical behaviour}
\label{sec.datacfgamma}
In this section, the prediction by a gamma distribution for two well-documented statistical properties in a plume is compared with the experimental data.
\subsection{Skewness and Kurtosis}
\label{ssec.skku}
For a gamma distribution, skewness and kurtosis are functions of the shape parameter and thus related as,
\begin{equation}
	\begin{rcases}
		\text{Skewness}\ \tilde{C}_\text{sk} = 2/\sqrt{k}\\
		\text{Kurtosis}\ \tilde{C}_\text{ku} = 3+6/k \\
	\end{rcases}
	\Rightarrow \tilde{C}_\text{ku} =1.5(\tilde{C}_\text{sk}^2)+3.
	\label{eq.SkKuGamma}
\end{equation}
Comparing with equation \eqref{eq.KuSk}, $A = 1.5$ and $B=3$ for the gamma distribution.
\begin{figure}
	\begin{center}
		\includegraphics[scale=1]{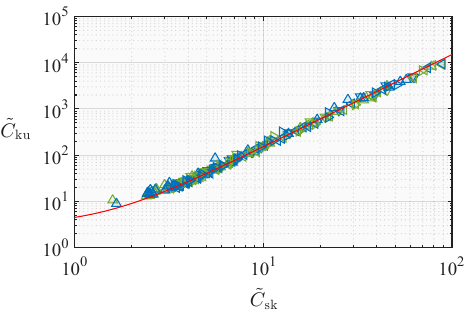}
	\end{center}
	\caption{Kurtosis versus skewness of concentration fluctuations. Solid red line: Equation \eqref{eq.SkKuGamma}. Symbols as defined in table \ref{tb.config}. }
	\label{fig.sk_vs_ku}
\end{figure}
The kurtosis $\tilde{C}_\text{ku}$ is plotted against the skewness $\tilde{C}_\text{sk}$ for all measurements in  figure \ref{fig.sk_vs_ku}. The trend of the data is consistent with that observed in past experimental studies \citep{Mole1995_CHighMoments, Schopflocher2005_SkKu, Nironi2015_P1}. For the present data, equation \eqref{eq.SkKuGamma} is in very good agreement with the data throughout the range except for a few outliers. It should be noted that both skewness and kurtosis increase with distance away from the centreline, and as such, the edge of the plume has high $\tilde{C}_\text{ku}$ and $\tilde{C}_\text{sk}$; this being consistent with the shape parameter ($k$) decreasing with distance from the plume centreline as observed in figure  \ref{fig.GammaFitParameters}(b). Figure \ref{fig.sk_vs_ku} conclusively shows that the predicted $\tilde{C}_\text{ku}$ versus ${C}_\text{sk}$ relationship by the gamma distribution is valid across the whole plume. Independent prediction of skewness and kurtosis is presented in $\S$\,\ref{sec.cn}.
\par The experimentally observed $\tilde{C}_\text{ku}$\,vs\,${C}_\text{sk}$ relationship in figure \ref{fig.sk_vs_ku} also supports the idea of statistical similarity of high concentration events as discussed in $\S$\,\ref{ssec.premultipliedpdf}. Other PDF distributions considered in the literature do not reproduce the experimental trends between skewness and kurtosis. Although skewness and kurtosis can be related for a Weibull and Poisson distributions, they differ markedly from the experimental data in this study. For example, for the Poisson distribution, $\tilde{C}_\text{ku}=3-5/\tilde{C}_\text{sk}^2$. For the generalised Pareto distribution (GPD), kurtosis is not a function of skewness, and no one-to-one relationship exists between the two. In the case of a normal or exponential distribution, kurtosis has a constant value and does not vary with skewness.
\subsection{The 99th percentile of instantaneous concentration}
\label{ssec.c99}
\begin{figure}
	\includegraphics[scale = 1]{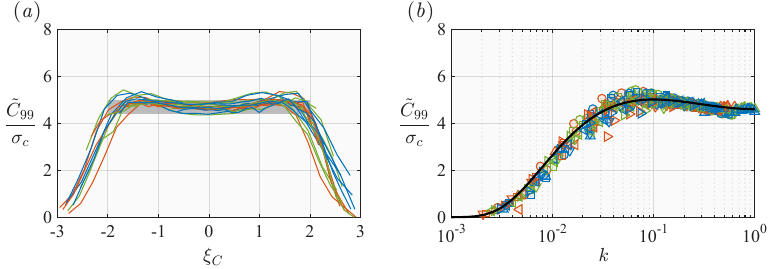}
	\caption{(\textit{a}) Ratio between the 99\textsuperscript{th} percentile of instantaneous concentration and the RMS, $\tilde{C}_{99}/\sigma_c$. (\textit{b}) $\tilde{C}_{99}/\sigma_c$ vs. $k$ from the experiments compared with theory. The black solid line is calculated numerically for the gamma distribution. Symbols as defined in table \ref{tb.config}. }
	\label{fig.c99}
\end{figure}
Another quantity of interest, examined in earlier studies, is the ratio between the 99\textsuperscript{th} percentile of the instantaneous concentration and the RMS, $\tilde{C}_{99}/\sigma_c$. Originally reported empirically by \citet{Fackrell1982}, this ratio is particularly relevant for regulatory applications as it provides a statistical basis for defining concentration thresholds. Figure \ref{fig.c99}(\textit{a}) shows the profiles of $\tilde{C}_{99}/\sigma_c$ as a function of $\xi_C$. Within $|\xi_C|\lessapprox2$ of the plume centreline, the ratio remains nearly constant, ranging between 4.2 and 4.9, with a local minimum at the centreline. For $|\xi_C|>2$, however, $\tilde{C}_{99}/\sigma_c$ decreases rapidly toward zero, reflecting diffusion and entrainment at the interface between the plume and the ambient fluid. \cite{Csanady1967_cfluc} showed that $\tilde{C}_{99}/\sigma_c\approx4.6$ under the assumption of a Poisson distribution. Experimental studies by \cite{Fackrell1982} reported values ranging between 4.5 and 5.0, while \cite{Lim2023_plume_piv} obtained a mean value of 4.6. These literature values align well with our present findings in the region, $|\xi_C|<2$.
\par For a gamma distribution, the ratio $\tilde{C}_{99}/\sigma_c$ can be obtained by evaluating the incomplete gamma function at the point where the cumulative density function (CDF) equals 0.99. With an appropriate change of variables, this yields:
\begin{equation}
	0.99 = \int_0^{\tilde{C}_{99}/\sigma_c}\frac{k^{k/2}}{\Gamma(k)}\,s^{k-1}\,\exp\left( -\sqrt{k}\,s \right) \mathrm{d}s
	\label{eq.C99}
\end{equation}
where $s$ represents a random variable. We note that the above expression is independent of the scale parameter $\theta$. 
Hence for a given $k$, one can determine a unique ratio $\tilde{C}_{99}/\sigma_c$. Accordingly, figure \ref{fig.c99}(\textit{b}) compares the measured values of $\tilde{C}_{99}/\sigma_c$ with those predicted from the fitted $k$, using the numerical relationship in equation \eqref{eq.C99}. The numerically estimated ratio shows excellent agreement with the experimental data, yielding $\tilde{C}_{99}/\sigma_c \approx 4.5$–$5$ near the plume centreline and remaining consistent across the entire plume. Remarkably, the gamma distribution captures the behaviour even at the plume edges, accurately predicting the much lower values of $\tilde{C}_{99}/\sigma_c$. \citet{Lim2023_plume_piv} recently proposed that $\tilde{C}_{99}/\sigma_c \sim 4.6$ is a robust and universal ratio, independent of intermittency levels or streamwise and vertical positions. Our results partially support this claim: the values of $\tilde{C}_{99}/\sigma_c$ are indeed robust, showing little sensitivity to source distance or source height in the present experiments. However, we find that the ratio is not universal, as it exhibits a systematic variation across the plume, albeit weak within $|\xi_C| \lesssim 2$.

\par Equation \eqref{eq.C99} is not only of theoretical interest but also of practical significance. The close agreement between the measured and predicted values of $\tilde{C}_{99}/\sigma_c$ demonstrates that, once the mean and variance of a concentration signal are known (for example, from RANS or LES simulations), assuming a gamma distribution and determining $k$ from equation \eqref{eq.GammaK} provides a reliable means of estimating concentration levels associated with specific risk thresholds.
\section{Higher-order moments}
\label{sec.cn}
\begin{figure}
	\centering
	\includegraphics[scale=1]{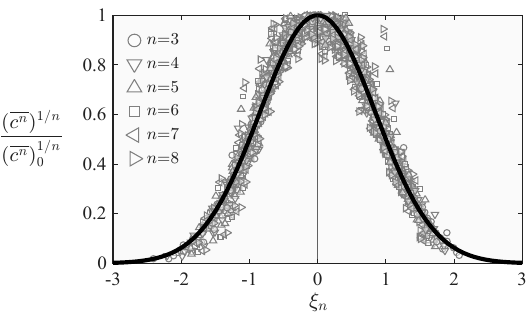}
	\caption{Normalised profiles of $(\overline{c^n})^{1/n}$ compared with the Gaussian profile (solid black line) for $n$ = 3 \textendash\ 8.}
	\label{fig.NormalisedMoments}
\end{figure}
In this section, the higher-order moments of concentration fluctuations are examined for both the experimental data and the gamma distribution. For the experimental data, the $n$-th order central moment is computed as:
\begin{equation}
	\overline{c^n}= \overline{(\tilde{C}-C)^n}, \;n>1.
	\label{eq.centralmoment}
\end{equation}
Likewise, the normalised $n$th-order central moment is defined as $\overline{c^n}/\overline{c^n_0}$, where $\overline{c^n_0}$ denotes its centreline value. In analogy with the mean and RMS distributions shown in figures \ref{fig.GaussianMeanStdDev}(\emph{a},\emph{b}), the normalised central moments are plotted against the normalised wall-normal coordinate in figure \ref{fig.NormalisedMoments} for $n=3$ to $8$. As with the mean and RMS, the moments decrease with distance from the plume centreline, allowing a half-width $\delta_n$ to be defined at the location where $(\overline{c^n})^{1/n}$ falls to half its centreline value. This gives the non-dimensional coordinate $\xi_n = (z-z_0)/\delta_n$. When plotted in this form, the profiles of the normalised moments exhibit Gaussian behaviour. Upon rescaling with the three parameters of equations \eqref{eq.CmeanGauss}—namely, the centreline height, centreline magnitude, and half-width—the profiles collapse onto a single curve \citep[as also shown by][]{Yee2000_salineInH2O}. Overall, the Gaussian model provides a good representation of the data, although figure \ref{fig.NormalisedMoments} shows increased scatter compared with figure \ref{fig.GaussianMeanStdDev}, reflecting the longer sampling required for statistical convergence of higher order moments \citep{Sreenivasan1980_tempfluc, Chatwin1990_CCrms}. A general Gaussian expression for the $n$th-order central moment is therefore given as,
\begin{equation}
	\frac{\overline{c^n}(z)}{\overline{c^n}(z=z_0)}= \exp\left[-(\ln2)\cdot n\cdot \xi_n^2\right], \text{ where } \xi_n = \frac{z-z_{\text{0}}}{\delta_{n}}\text{ and } \  n>1.
	\label{eq.cn_gauss}
\end{equation}
Past experimental studies \citep[e.g.][]{Nironi2015_P1, Yee2000_salineInH2O} have shown that skewness and kurtosis exhibit Gaussian, bell-shaped profiles with maxima at the plume centreline. In particular, \cite{Yee2000_salineInH2O} employed the normalisation $\overline{c^n}/C_0^n$ and reported only skewness and kurtosis, which are the standardised central moments. In contrast, the present analysis demonstrates Gaussian similarity across the normalised central moments—analogous to the mean and RMS—through equation \eqref{eq.cn_gauss}, extending up to $n=8$. To our knowledge, no previous studies have validated a probability distribution against experimental data to this degree. 

\par
\begin{figure}
	\centering
	\includegraphics[scale = 1]{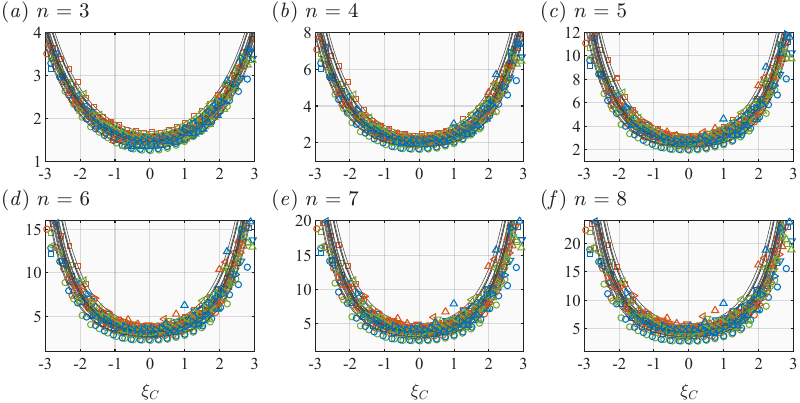}
	\caption{Profiles of $(\overline{c^n}/\sigma_c^n)^{1/n}$ compared with the predictions from gamma distribution (solid lines) for $n$ = 3 \textendash\ 8 estimated using equation \eqref{eq.cn_prediction}. Symbols as defined in table \ref{tb.config}.}
	\label{fig.StandardisedMoments}
\end{figure}

 It is therefore natural to examine the standardised central moments also. By definition, the $n$th-order standardised central moment is obtained by dividing the central moment (equation \eqref{eq.centralmoment}) by the standard deviation raised to the $n$th power:
\begin{equation}
	\frac{\overline{c^n}}{\sigma_c^{n}}= \frac{\overline{(\tilde{C}-C)^n}}{\sigma_c^n},\; n>1
	\label{eq.standardmoment}
\end{equation}
For $n$ = 3 and 4, the standardised central moments are skewness and kurtosis, respectively. Since $\tilde{C}\ge 0$ and $\mathcal{P}(\tilde{C}=0)\ne 0$ (due to intermittency), the distribution is positively-skewed and thus all central moments are positive too. The standardised central moments are plotted in figure \ref{fig.StandardisedMoments}, again raised to the exponent $1/n$ to have a similar order of magnitude. The magnitude of standardised moments becomes significantly large at $|\xi_C| \gtrsim 2$, consistent with observations by \cite{Chatwin1990_CCrms} and \cite{karnik1989heatdiffusion} for skewness. This is because of the scalar intermittency (durations of $\tilde{C}>0$), which decreases away from the centreline. Also, within $|\xi_C| \lesssim 2$, the maximum instantaneous concentration is significantly higher than the mean concentration, that is $\tilde{C}_\mathrm{max} \gg C$.
\par It should be noted that the standardised central moments in figure \ref{fig.StandardisedMoments} are plotted against the normalised distance from the centreline, $\xi_C$, i.e. based on the mean concentration profile. This is deliberate, as the standardised moments can be expressed in terms of the mean concentration profile $C(z)$ or as a function of $\xi_C$, such that $\xi_C$ and $\xi_n$ are related. A brief derivation is provided here. For the two-parameter gamma distribution, the expected (mean) \emph{raw} moments (relative to 0 or absolute magnitude) are given as,
\begin{equation}
	\mathbb{E}(\ci^n) = \theta^n\frac{\Gamma(k+n)}{\Gamma(k)} = \frac{C^n}{k^n}\frac{\Gamma(k+n)}{\Gamma(k)}.
\end{equation}
Here, $\mathbb{E}(\ci^n)$ is the expected value of the $n$th raw moment and $\Gamma(k+n) = (k+n-1)!$ and $\Gamma(k) = (k-1)!$ are the gamma function or the factorial function. The raw moments are not used to study dispersion, but they are useful for deriving the central moments. Accordingly, the central moments (relative to the mean) are
\begin{equation}
	\overline{c^n} = \mathbb{E} \left[ (\tilde{C}-C)^n\right] =C^n \cdot \sum_{m=1}^n a_m^n k^{m-n}, 
\end{equation}
where $a_m^n$ are the binomial coefficients that are specific to each $n$ and can be found in Appendix \ref{appen.deduction} for $n$ upto 8. Substituting $\sigma_c = C/\sqrt{k}$ (from equations \eqref{eq.GammaMean} and \eqref{eq.GammaCrms}), the standardised central moment can be described by the generalised form,
\begin{equation}
	\frac{\overline{c^n}}{\sigma_c^n} = \sum_{m=1}^n a_m^n k^{m-n/2}. \label{eq.cn_standardised}
\end{equation}
Further substitution of equation \eqref{eq.kGauss} into equation \eqref{eq.cn_standardised} gives the standardised moments as a function of the distance from the centreline, $\xi_C$:
\begin{equation}
	\frac{\overline{c^n}}{\sigma_c^n} = \sum_{m=1}^n \left(a_m^n k_0^{m-n/2}\exp\left[-(\ln 2)\left(m-n/2\right)\xi_C^2\right]\right). \label{eq.cn_prediction}
\end{equation}
Accordingly, if $\delta_C$ and $k_0$ are known, then the higher order moments can be predicted at any height $z$ relative to the centreline. The two methods for determining $k_0$ were presented in $\S$\,\ref{ssec.gammavalidation}. Equation \eqref{eq.cn_prediction} is the main result of this section and is used to predict the higher-order moments in figure \ref{fig.StandardisedMoments} as the solid lines. The predicted values are in good agreement with the experimental data, especially for $n=3$ -- $6$. There is increased scatter in the experimental data for $n=7, 8$, but the predictions still capture the general trend. The applicability of equation \eqref{eq.cn_prediction} is limited to far-field regions of a plume whose width is thin relative to the boundary layer thickness. In the near-field, concentration distributions can be Gaussian, and centreline skewness is negative \citep{Sawford_1995_moments, Cassiani2024_cfluc_LES}. 
\par
\begin{figure}
	\centering
	\includegraphics[width = 0.49\textwidth]{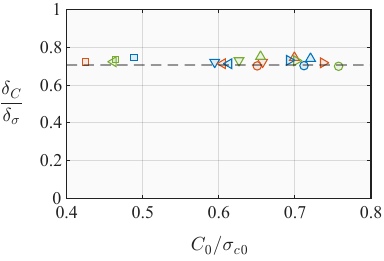} 
	\caption{The ratio of the half-plume width from the mean profiles to the half-plume width from the RMS profiles vs. the ratio between the centreline values of mean and RMS. The horizontal dashed line represents $\delta_C/\delta_\sigma=1/\sqrt{2}$. Symbols as defined in table \ref{tb.config}.}
	\label{fig.deltaratio}
\end{figure}

 Alternatively, using equation \eqref{eq.cn_gauss} and the exponentiation rule for the Gaussian term,

\begin{align}
	\frac{\overline{c^n}\left(z\right)}{\big(\overline{c^n}\big)_0}
	& =  \exp\left[-(\ln2)\cdot n\cdot \xi_n^2\right] = \frac{g(k,n)}{g(k_0,n)}\cdot \left(\frac{C}{C_{0}}\right)^n\\
	\therefore & \exp\left[-(\ln2)\cdot n \cdot \left(\frac{\delta_
	C^2}{\delta_n^2}-1\right) \cdot \xi_C^2\right] = \frac{g(k,n)}{g(k_0,n)}
\end{align}
Here $g(k,n)$ is a polynomial in $k$. Since $k$ follows the same Gaussian distribution as $C$, as discussed in $\S$ \ref{ssec.gammavalidation}, when $\xi_C=1$, $k = k_0/2$. For $n=2$, $g(k,n=2) = 1/k$. Therefore,

\begin{align}
	\exp\left[(\ln2)\cdot n \cdot \left(1-\frac{\delta_
	C^2}{\delta_n^2}\right)\right] = 2\quad \Rightarrow \quad \frac{\delta_C}{\delta_n} = \left(1-\frac{1}{n}\right)^{1/2} \label{eq.5.10}
\end{align}
For $n=2$, $\delta_2 = \delta_\sigma$; and thus $\delta_C/\delta_\sigma = 1/\sqrt{2}$. This identity is also a novel contribution of this paper and is validated by the experimental results in figure \ref{fig.deltaratio}. Further, Appendix \ref{append.delta_ratio} provides arguments for the consistency of this relationship with the data for the shape and scale parameters.
\par To summarise this section, a formulation for standardised central moments is provided and validated against the experimental data. The formulation relies on \emph{a priori} information of $C(z)$ and $\sigma_{c,0}$ to determine the statistical behaviour within the entire plume, under the assumption that the gamma distribution is a valid probability distribution for instantaneous events. This assumption has been thoroughly validated in the present and previous sections.
\section{Discussion}
\label{sec.discussion}
\par Several aspects in the previous sections merit further discussion. It is shown in $\S$\ref{ssec.gammavalidation} that the decay of exponential tails remains approximately constant within the core region, $-1\!\lesssim \xi_C \lesssim 1$. This finding extends theoretical predictions by \cite{Pumir1991_PDF_expTail, 1994_Shraiman_PDF_exponentialTail} in support of the universality argument by \cite{Mole2008}, suggesting that the exponential decay is governed by uniform physical processes across the core of a slender plume. It is hypothesised that the constant $\theta$ arises from the interplay between small-scale mixing and large-scale stirring mechanisms within the plume. The small-scale mixing, driven by molecular diffusion, acts to homogenise scalar concentrations, resulting in a decrease of scalar magnitude. However, the large-scale stirring, caused by energetic turbulent structures, redistributes scalar concentrations throughout the plume while maintaining the scalar magnitude; i.e., the process is dictated by advection. This interpretation is consistent with the self-similar behaviour of the scalar spectra documented by \cite{Talluru2019_Spectra}, where equal distribution of \emph{scalar energy} is found across the plume. For a slender plume, the transverse mixing by $v$ and $w$ fluctuations would result in such a distribution of instantaneous scalar fluctuation if the integral length-scales associated with these fluctuations are of the same order as the plume half-width. In the streamwise direction, the plume is primarily advected. Thus, the background turbulent flow, responsible for self-similar scalar spectra, also contributes to the constant $\theta$ across the plume. Further investigation, preferably using all three velocity components along with scalar, is required to describe the underlying phenomena.

\par Secondly, the range of Schmidt number in this study is within 0.29 -- 0.78 (for varying density ratios), typical for wind tunnel experiments. Plumes with a higher Schmidt number would undergo relatively less homogenisation by molecular diffusion. Experiments conducted in water channels have $\mathcal{O}(Sc)\sim 10^3$ \citep{Yee1993b, 2011_Skvortsov_cn_waterchannel, Yee2011_gammaPDF, Vanderwel2014}, and have reported the gamma distribution for instantaneous concentration. However, the distribution of $k$ or $\theta$ across the plume is not reported in these studies.

\par It is possible that buoyancy in a heavy or light plume would have a substantial influence downstream, not only on the scalar statistics but also on instantaneous behaviour (such as profiles of $k$ and $\theta$). However, we observe that the invariant characteristics of buoyant plumes are similar to those of a neutral plume. The invariance can be attributed to the turbulence generation mechanism in buoyant plumes; as elucidated by \cite{1994_Shabbir_buoyantPlume_turbulenceProduction} to state ``the primary effect of buoyancy on turbulence is indirect, and enters through the mean velocity field''.

Further, the Richardson numbers achieved at the source in our study are still relatively low, and even lower at the downstream measurement locations. Thus, buoyancy effects are limited to the orientation of the plume in our study, such that they vary the plume height and the half-plume width. For the same source height and measurement distance, positively buoyant plumes achieve a higher centreline position and wider half-widths, whereas negatively buoyant plumes have the opposite behaviour, relative to the neutral plume \citep{Pang2024_plume_exp}. Similarly, whilst $k$ and $\theta$ values vary with source buoyancy, the Gaussian behaviour of $k(z)$ and constant decay exponent of the PDF tail remain invariant. Therefore, equation \eqref{eq.cn_prediction} is valid irrespective of whether the plume is positively, neutrally, or negatively buoyant. Thus, at least for $Ri\ll1$ in the meandering regime, it is concluded that high concentration \emph{parcels} are still transported by the background turbulence, and thus act as passive scalars.

\par The findings of the present study can also be implemented as a robust benchmark for numerical solution of the time-dependent scalar-transport equation coupled with Reynolds-averaged Navier-Stokes or Large-Eddy Simulation methods. Eddy-diffusivity formulations or sub-grid scale models can be assessed in simple flow geometries for their ability to predict scalar fluctuations that align with the observations in sections \ref{sec.datacfgamma} and \ref{sec.cn}.

\par The current study focuses on elevated point sources in a turbulent boundary layer. Future research should explore the applicability of the gamma distribution and equation \ref{eq.cn_prediction} to other configurations, such as ground-level releases, area sources, and different atmospheric stability conditions. For the case of ground-level source, it is known that the background near-wall mean flow and turbulence have higher shear and anisotropy, respectively, whereas, in the outer region of the turbulent boundary layer, the fluctuations (in wall-normal and transverse directions) are well-represented by the Gaussian PDF, with near-zero skewness and excess kurtosis \citep{Baidya2016_thesis}. Despite these differences, the gamma distribution is frequently evidenced for scalar fluctuations in plume emerging from a ground-level source \citep{Yee2011_gammaPDF, Nironi2013_thesis, Lim2023_plume_piv}. Thus, extending these observations to the variation of shape and scale parameters in these flows, akin to this study, is of interest.
\section{Conclusion}
The validity of the gamma distribution for describing concentration PDFs in scalar plumes in a turbulent boundary layer is comprehensively assessed using high-quality experimental data. The distribution demonstrates excellent agreement with experimental observations across multiple theoretical properties, including previously unexplored characteristics. Most notably, we demonstrate for the first time that the gamma distribution accurately predicts the ratio of the 99$^\text{th}$ percentile to RMS concentration across the entire plume width, from core to edge. The two parameters of the gamma distribution are fitted to the data. The shape parameter exhibits a Gaussian-like profile, while the scale parameter remains approximately constant across the plume, confirming the constant exponential decay of PDF tails in a plume, which is not captured in other distributions. Both parameters agree well with theoretical predictions derived from the gamma distribution solely using the mean and the RMS concentration. Building on these validated relationships, we propose a novel method for predicting higher-order concentration moments using the gamma distribution framework. This method requires only the mean and RMS concentration as inputs, eliminating the need for additional parameterisation required by previous methods, which have been verified up to the 8th-order moment.

\appendix
\section{Distributions}
\label{appen.dist}
  The expressions for the PDFs are listed in table \ref{tb.litrev.PDFs}. Among the listed distributions, normal, beta, log-normal, and gamma distributions belong to the exponential family. The Weibull distribution can have an exponential tail when the parameter $k$ is fixed.
\begin{table}
	\centering
	\begin{tabular}{p{0.2\textwidth}p{0.55\textwidth}l} 
		\toprule
		Name & Function & Parameters\\
		\midrule
		Normal & $\mathcal{P}(x)=\frac{1}{\sigma \sqrt{2\pi}}\exp\left[ {-\frac{1}{2}\left( \frac{x-\mu}{\sigma}\right)^2}\right]$ & $\mu$, $\sigma$\medskip\\
		Log-normal & $\mathcal{P}(x)=\frac{1}{x\sigma \sqrt{2\pi}}\exp\left[{-\frac{1}{2}\left( \frac{\ln{x}-\mu}{\sigma}\right)^2}\right]$ & $\mu$, $\sigma$\medskip\\
		Beta & $\mathcal{P}(x)=\frac{x^{\alpha-1}(1-x)^{\beta-1}}{B(\alpha,\beta)}$, where $B(\alpha,\beta)=\frac{\Gamma(\alpha)\Gamma(\beta)}{\Gamma(\alpha+\beta)}$ & $\alpha$, $\beta$\medskip\\
		Gamma & $\mathcal{P}(x)=\frac{1}{\Gamma(k)\theta^k}x^{k-1}\exp\left(-\frac{x}{\theta}\right)$ & $k$, $\theta$\medskip\\
		Weibull & $\mathcal{P}(x)=\frac{k}{\lambda}\left( \frac{x}{\lambda} \right)^{k-1}\exp\left[-\left(\frac{x}{\lambda}\right)^k\right]$ & $k$, $\lambda$\medskip\\
		Generalised Pareto & $\mathcal{P}(x) = \frac{1}{\sigma}\left(1+\xi \frac{x-\mu}{\sigma}\right)^{-1/\xi+1}$ & $\mu$, $\sigma$\\ \bottomrule
	\end{tabular}
	\caption{The probability density functions proposed in past literature to describe the instantaneous concentration. Note that the $x$ in this table is a random variable, different from the $x$ defined as streamwise location in the rest of this manuscript.}
	\label{tb.litrev.PDFs}
\end{table}
\section{Vertical profile of gamma parameters: $k$ and $\theta$}
\label{append.delta_ratio}
It is reasonable to consider that the plume centreline height for all moments is the same for elevated sources. Consider the ratio of half-widths,

\begin{equation*}
	\frac{\delta_C}{\delta_\sigma}=\sqrt{D},
\end{equation*}

It follows from equations \eqref{eq.CmeanGauss} and \eqref{eq.CrmsGauss},
\begin{equation}
	\frac{\sigma_c(z)}{\sigma_{c, 0}} =\left(\frac{C(z)}{C_{0}}\right)^D
	\label{eq.appenB5}
\end{equation}
Multiplying both sides by $C_{0}/C(z)$ and rearranging,
\begin{equation}
	\frac{\sigma_c(z)}{C(z)} \frac{C_{0}}{\sigma_{c,0}}=\left(\frac{C(z)}{C_{0}}\right)^D\frac{C_{0}}{C}
	\label{eq.appen.step2}
\end{equation}
\textbf{Assumption:} The gamma distribution is the appropriate PDF for the instantaneous concentration.

\begin{align*}
	& C(z) = k(z)\theta,\\
	& C_0 = k_{0}\theta_{0}, \\
	& \sigma_c(z) = \sqrt{k(z)}\theta ,\\
	& \sigma_{c,0} = \sqrt{k_{0}}\theta_{0}.
\end{align*}
Substituting the above expressions into equation \eqref{eq.appen.step2} and rearranging,
\begin{equation}
	\frac{k(z)}{k_0} = \left(\frac{C(z)}{C_0}\right)^{2-2D}
	= \left[ \exp  \left( -(\ln2)\;\xi_C^2\right) \right]^{2-2D}.
	\label{eq.appen.kgauss}
\end{equation}
Hence, the vertical profile of $k$ is Gaussian. Experimental data in figure \ref{fig.GammaFitParameters}(\textit{b}) verifies the \textbf{assumption} made regarding the gamma distribution.
\par Similarly, for $\theta(z)/\theta_0$, taking the square of equation \eqref{eq.appenB5} and rearranging leads to,
\begin{equation}
	\frac{\theta(z)}{\theta_0} = \left( \frac{C(z)}{C_0} \right)^{2D-1}
	\label{eq.appen.thetagauss}
\end{equation}
We have observed in figure \ref{fig.GammaFitParameters}(\textit{d}) that $\theta(z)$ does not vary significantly from its centreline value over $|\xi_C|\le 1$. This behaviour is only achieved for equation \eqref{eq.appen.thetagauss} if $D = 1/2$.
\par Alternatively, one can start from the experimental evidence in figure \ref{fig.deltaratio} that $\delta_C/\delta_\sigma=1/\sqrt{2}$, meaning $D=1/2$. Substituting the value of $D$ into equations \eqref{eq.appen.kgauss}, one gets
\begin{equation}
	\frac{k(z)}{k_0} = \exp  \left[ -(\ln2)\;\xi_C^2\right]
\end{equation}
Thus, the Gaussian variation of $k(z)$ with $\xi_C$ is confirmed. It is also possible to first note that in figure \ref{fig.GammaFitParameters}(\textit{b}) that the half-width of $k$ profiles is the same as that of the $C$ profiles, i.e. $k=k_0/2$ at the same $z$ where $C=C_0/2$. This means that the exponent $(2-2D)$ in equation \eqref{eq.appen.kgauss} is equal to one.

\begin{equation*}
	2-2D=1\; \Rightarrow\; D=1/2
\end{equation*}
Substituting the value of $D=1/2$ to equation \eqref{eq.appen.thetagauss}, again gives,

\begin{equation*}
	\frac{\theta(z)}{\theta_0} = 1,
\end{equation*}
and is consistent with the data in figure \ref{fig.GammaFitParameters}(\textit{d}). Thus, $\theta$ = constant is consistent with the gamma distribution being the appropriate distribution for the plume.
\section{Equations for $\mathbb{E}({\tilde{C}^n})$ and $\overline{c^n}$ for the gamma distribution}
\label{appen.deduction}
Adopting the gamma distribution as the appropriate distribution for concentration fluctuations, the $n$th raw moment about zero can be expressed in terms of $k$ and $\theta$ as,

\begin{align}
	\mathbb{E}(\ci^n) & = \theta^n\frac{\Gamma(k+n)}{\Gamma(k)} = \frac{C^n}{k^n}\frac{\Gamma(k+n)}{\Gamma(k)}.
\end{align}
For the first eight raw moments, these can be evaluated from the following:

\begin{align*}
	\begin{bmatrix}
		\mathbb{E}(\tilde{C}^2)/C^2\\
		\mathbb{E}(\tilde{C}^3)/C^3\\
		\mathbb{E}(\tilde{C}^4)/C^4\\
		\mathbb{E}(\tilde{C}^5)/C^5\\
		\mathbb{E}(\tilde{C}^6)/C^6\\
		\mathbb{E}(\tilde{C}^7)/C^7\\
		\mathbb{E}(\tilde{C}^8)/C^8\\
	\end{bmatrix}
	=
	\begin{bmatrix}
		1 & 1 & 0 & 0 & 0 & 0 & 0 & 0\\
		1 & 3 & 2 & 0 & 0 & 0 & 0 & 0\\
		1 & 6 & 11 & 6 & 0 & 0 & 0 & 0\\
		1 & 10 & 35 & 50 & 24 & 0 & 0 & 0\\
		1 & 15 & 85 & 225 & 274 & 120 & 0 & 0\\
		1 & 21 & 175 & 735 & 1624 & 1764 & 720 & 0\\
		1 & 28 & 322 & 1960 & 6769 & 13132 & 13068 & 5040\\
	\end{bmatrix}
	\times
	\begin{bmatrix}
		1/k^0\\
		1/k^1\\
		1/k^2\\
		1/k^3\\
		1/k^4\\
		1/k^5\\
		1/k^6\\
	\end{bmatrix}
\end{align*}

The $n$th moment about the mean can therefore be expressed as

\begin{align}
	\overline{c^n} & = \mathbb{E} \left[ (\tilde{C}-C)^n\right] = \theta^n\sum_{m=1}^n a_m k^m
\end{align}
For the first eight central moments, these can be evaluated from the following:

\begin{align*}
	\begin{bmatrix}
		\overline{c^2}/\theta^2\\
		\overline{c^3}/\theta^3\\
		\overline{c^4}/\theta^4\\
		\overline{c^5}/\theta^5\\
		\overline{c^6}/\theta^6\\
		\overline{c^7}/\theta^7\\
		\overline{c^8}/\theta^8\\
	\end{bmatrix}
	=
	\begin{bmatrix}
		1 & 0 & 0 & 0 & 0 & 0\\
		2 & 0 & 0 & 0 & 0 & 0\\
		3 & 6 & 0 & 0 & 0 & 0\\
		0 & 20 & 24 & 0 & 0 & 0\\
		0 & 15 & 130 & 120 & 0 & 0\\
		0 & 0 & 210 & 924 & 720 & 0\\
		0 & 0 & 105 & 2380 & 7308 & 5040\\
	\end{bmatrix}
	\times
	\begin{bmatrix}
		k\\
		k^2\\
		k^3\\
		k^4\\
		k^5\\
		k^6\\
		k^7\\
	\end{bmatrix}
\end{align*}
The standardised moment is only a function of $k$,

\begin{align}
	\frac{\overline{c^n}}{(\overline{c^2})^{n/2}} & = \frac{\sum_{m=1}^n a_mk^m}{k^{n/2}}.
\end{align}
\clearpage

\bibliographystyle{unsrtnat}

\bibliography{jfm}

\end{document}